\documentstyle[aps,preprint]{revtex}
\newcommand{\be}{\begin{equation}}
\newcommand{\ee}{\end{equation}}
\newcommand{\bea}{\begin{eqnarray}}
\newcommand{\eea}{\end{eqnarray}}

\input epsf

\begin{document}
\pagestyle{plain}
\bibliographystyle{prsty}
\title{Tunable front interaction and localization of periodically forced waves}
\author{Catherine Crawford\thanks{Corresponding author.
Current Address: Elmhurst College, Department of Mathematics, 190
Prospect Ave., Elmhurst, IL 60126; Tel: (630) 617-3479; E-mail:
crawford@elmhurst.edu} and Hermann Riecke}
\address{Department of Engineering Sciences and Applied Mathematics,
Northwestern University, Evanston, IL 60208, USA}
\date{\today}
\maketitle

\begin{abstract}

In systems that exhibit a bistability between nonlinear traveling waves and the
basic state, pairs of fronts connecting these two states can form localized
wave pulses whose stability depends on the interaction between the fronts.  We
investigate wave pulses within the framework of coupled
Ginzburg-Landau equations describing the traveling-wave amplitudes. We find
that the introduction of resonant temporal forcing  results in a {\em new,
tunable} mechanism for stabilizing such wave pulses. In contrast to other
localization mechanisms the temporal forcing can achieve localization by a
repulsive as well as by an attractive interaction between the fronts. Systems
for which the results are expected to be relevant include binary-mixture
convection and electroconvection in nematic liquid crystals.

\end{abstract}

\section{Introduction}

Localized structures have been observed in a range of pattern-forming
nonequilibrium systems. One type of localized structure occurs when one pattern
is embedded within another pattern. Examples include the coexisting stationary
domains of long and short wavelengths observed in Taylor-Couette flow between
co-rotating cylinders \cite{BaAn86}, in Rayleigh-B\'{e}nard convection in
narrow slots \cite{HeVi92}, and in parametrically excited waves in ferrofluids
\cite{MaRe98}.  Solitary waves drifting through a stationary pattern are found
associated with a parity-breaking bifurcation in directional solidification
\cite{SiBe88}, the printer instability \cite{RaMi90}, viscous fingering
\cite{CuFo93}, cellular flames \cite{GoEl94}, and Taylor-vortex flow
\cite{WiMc92}.

In this paper, we investigate a class of localized states in which the
pattern is confined to a small region that is surrounded by the unpatterned state, or
vice versa.  For example, solitary standing waves (`oscillons') have been
observed in vertically vibrated granular layers \cite{UmMe96} and colloidal
suspensions \cite{LiHa99}.   Localized traveling waves have been observed as
one-dimensional pulses in binary-fluid mixtures
\cite{MoFi87,KoBe88,NiAh90,Ko91,Ko94} and as two-dimensional `worms' in
electroconvection of nematic liquid crystals \cite{DeAh96a}.

For a general understanding of such structures the mechanisms that are
responsible for their localization are of particular interest.  A number of
different types of localization mechanisms have been identified (e.g.
\cite{Ri99}).  To provide a context for our results we briefly review the
main mechanisms.

The stable coexistence of domains of long and short wavelengths can be
understood to be due to the instability of the constant wavenumber state
combined with the conservation of the phase
\cite{BrDe89,BrDe90,DeLe90,Ri90,RaRi95a}.  Localized patterns can also be
stabilized by a non-adiabatic pinning of the large-scale envelope to the
underlying small-scale pattern \cite{Po86,BeSh88}.  This pinning has been
suggested as a possible localization mechanism for oscillons \cite{CrRi99}.

To understand the traveling-wave pulses in binary mixture convection, two
mechanisms have been put forward, dispersion \cite{FaTh90,MaNe90,HaPo91} and
the advection of a slowly decaying concentration mode
\cite{Ri92,HeRi95,RiRa95,Ri96}.  Within the framework of the complex
Ginzburg-Landau equation with strong dispersion, pulses and holes can be viewed
as perturbed bright and dark solitons of the nonlinear Schr\"{o}dinger equation
\cite{ElMe89,ThFa88,LeFa97}.  For weak dispersion, pulses have been described
as a pair of bound fronts. In the absence of dispersion they are unstable.
However, dispersion may result in a repulsive interaction between the two
fronts and a stable pulse can arise \cite{MaNe90,HaPo91}.  Similarly, the
advected mode modifies the interaction between fronts and can also provide a
stabilizing repulsive interaction.  A similar advective mechanism has been
invoked \cite{RiGr98} to explain the two-dimensional localized waves (`worms')
that have been observed in electroconvection in nematic liquid crystals
\cite{DeAh96a}.

More generally, the coupling of a pattern to an additional undamped (or weakly damped) mode
can lead to its localization \cite{CoMa01}.  In the drift waves
arising from a parity-breaking bifurcation, the local wavenumber of the
underlying pattern plays the role of the additional mode \cite{CaCa92,RiPa92}.
For the oscillons in vibrated granular media it has been suggested that a
coupling of the surface wave to a mode representing the local height of the
granular layer is important \cite{TsAr97}.

For traveling waves it is well known that the external application of a resonant
temporal forcing excites the counter propagating wave
\cite{RiCr88,Wa88,RiCr90}. A natural question is therefore whether the
counterpropagating wave can play a role similar to the various additional modes
mentioned above and can thus lead to the localization of the traveling wave into a
pulse.  Since the temporal forcing is easily controlled externally this
localization mechanism would be {\em tunable}.

In this paper we investigate the effect of time-periodic forcing on
spatially localized waves that arise in systems exhibiting a subcritical
bifurcation to traveling waves as is, for instance, the case in binary-mixture
convection. We expect the results also to be relevant for the worms observed
in electroconvection in nematic liquid crystals.

We first consider the effect of forcing on the interaction of
fronts in the absence of other localization mechanisms and show
that forcing alone can lead to localized structures.  This new
localization mechanism can stabilize pulses with either a
repulsive or an attractive interaction.  While the interaction
strength between fronts is usually determined by the system
parameters, forcing allows the strength to be controlled
externally.  To focus on the interaction of the fronts due to the
temporal forcing, we start in Section \ref{sec:derivation} with
two coupled dispersionless Ginzburg-Landau equations and derive
evolution equations for the fronts.  In Section
\ref{sec:discussion} we discuss the resulting front equations and
compare them with numerical calculations. Section \ref{sec:holes}
extends the analysis and discussion to include holes and multiple
pulses.  The combined effect of temporal forcing and dispersion is
investigated in Section \ref{sec:disp} and conclusions are
presented in Section \ref{sec:conclusion}.

\section{Derivation of the front equations} \label{sec:derivation}

Motivated by the pulses observed in binary mixture convection
\cite{MoFi87,KoBe88,NiAh90,Ko91,Ko94} and the
worms in electroconvection \cite{DeAh96a} we consider a
subcritical bifurcation to traveling waves in a one-dimensional
system that is parametrically forced. To obtain a weakly nonlinear
description, physical quantities like the temperature ${\cal T}$ of the
fluid in the midplane in convection, say, are expanded in terms of the
amplitudes $A$ and $B$, of the left and right traveling waves,
\be
{\cal T} = \epsilon^{1/2}\left\{ A(x,t) e^{i(q_c\tilde{x}-\frac{\omega_e}{2} \tilde{t})} +
B(x,t)e^{-i(q_c\tilde{x}+\frac{\omega_e}{2} \tilde{t})} \right\} + c.c. + h.o.t.,
\qquad \epsilon \ll 1, \label{e:expand}
\ee
where $\tilde{t}$ and $\tilde{x}$ are the fast time and space
coordinates.  The amplitudes $A$ and $B$ are allowed to vary on a
slow time scale $t$ and a slow spatial scale $x$. Due to
the  external periodic forcing $\epsilon^2\,\nu e^{i \omega_e \tilde{t}}$ close to twice
the Hopf frequency,  $\omega_e = 2 (\omega_h+\epsilon^2\Omega)$,
the expansion (\ref{e:expand}) is chosen in terms of the forcing frequency.
The forcing excites the oppositely traveling waves and breaks the time
translation symmetry, resulting to lowest order in a linear
coupling between the two wave amplitudes.
 Using the remaining spatial
translation and reflection symmetries of the system,
 the form of the amplitude equations for $A$ and $B$ can
be derived \cite{RiCr88,Wa88,RiCr90}.
Hence we study the following set of
coupled Ginzburg-Landau equations as a model describing the system,
\bea
\partial_t A & = & -s\partial_x A + \epsilon^2 d_2 \partial_{xx} A
                   + \mu A +c|A|^2 A- p|A|^4A \nonumber \\
&&- g|B|^2 A - r|A|^2|B|^2 A -u|B|^4 A + \nu B^* \label{e:ampA} \\
\partial_t B & = & +s\partial_x B + \epsilon^2 d_2 \partial_{xx} B
                   + \mu B +c|B|^2 B -p|B|^4B \nonumber \\
&&- g|A|^2 B - r|A|^2|B|^2 B -u|A|^4 B + \nu A^*, \label{e:ampB}
\eea
where the forcing coefficient $\nu$ and the group velocity $s$ are
real. All other coefficients may be complex.  However,
to focus on the interaction of the fronts due to temporal forcing,
the front equations are derived for the case in which all of the
coefficients are real, i.e. neglecting dispersion and detuning.
Nonlinear gradient terms which would also appear in
 equations (\ref{e:ampA}) and (\ref{e:ampB}) to the order considered
have also been neglected.  In order for the bifurcation to be
weakly subcritical, the cubic coefficients must be small enough to allow a
balance with the quintic terms.

The bifurcation to spatially extended traveling waves in electroconvection in
nematic liquid crystals appears to be supercritical \cite{TrKr97}. To explain
the observation of worms already below the onset of spatially extended waves,
it has been argued that a weakly damped mode is relevant that is advected by
the wave \cite{RiGr98}. Already below threshold such a mode can lead to the
existence of infinitely long worms, i.e. to convection structures that are
narrow in the $y$-direction, say, and spatially periodic in the $x$-direction
\cite{RiGr98,Ri99}. They are expected to arise in a secondary bifurcation off
the spatially extended traveling waves \cite{RoRi01}. Focusing on the dynamics
of the worms in the $x$-direction, the head and the tail of the worms can be
considered as leading and trailing fronts that connect the nonlinear state,
which is strongly localized in the $y$-direction, with the basic non-convective
state. It is reasonable to expect that Ginzburg-Landau equations for a
subcritical bifurcation will capture the qualitative aspects of these
structures.

With the usual scaling $x=\epsilon \tilde{x}$
 complex Ginzburg-Landau equations are obtained in which the growth
term $\mu A$ is balanced by the diffusive term $d\partial_{xx} A$.
 For group velocities of order 1 this implies that the term $s\partial_{x} A$
 is inconsistent with the rest of the equation, i.e. it is of
{\em lower} order.  However, by considering slower spatial scales
$x = \epsilon^2 \tilde{x}$, the advective term $s\partial_x A$ is of the same
order as $\mu A$ and the diffusive term then appears in the
re-scaled equations (\ref{e:ampA}) and (\ref{e:ampB}) as a
higher-order correction of ${\cal O}(\epsilon^2)$ as indicated in
(\ref{e:ampA}) and (\ref{e:ampB}) \cite{MaVe96}.

\begin{figure}[ht]
\centerline{\epsfxsize=3.3in{\epsfbox{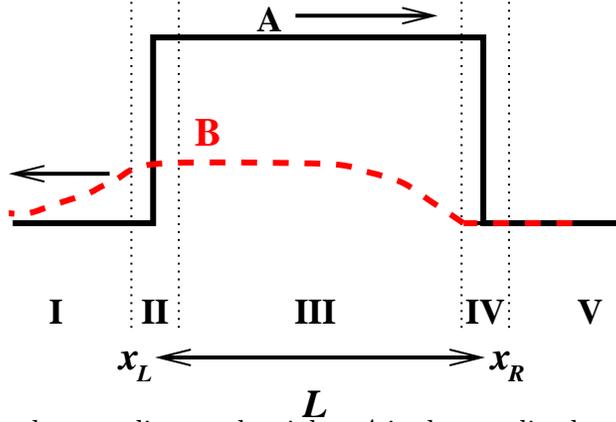}}
}
\caption{Sketch of a pulse traveling to the right.  $A$ is the amplitude of
the right-traveling wave.  Temporal forcing excites the left-traveling wave $B$,
which grows spatially to the left reaching a maximum at the trailing front,
behind which it decays exponentially.
        }
\label{fig:front}
\end{figure}

We are interested in localized solutions made up of two bound fronts
connecting the basic state with the nonlinear state as sketched in
Figure~\ref{fig:front}. One contribution to the interaction between the
fronts arises from their overlap.  For large distances it is small,
inducing a weak {\it attractive} interaction, which destabilizes pulses.
However, since the diffusion is weak (${\cal O}(\epsilon^2)$),
resulting in steep fronts in A, the overlap between the fronts is of
higher order in $\epsilon$ and the associated interaction can be
ignored.  Hence the interaction between the fronts is dominated by
the presence of the oppositely traveling wave excited by the periodic
forcing. The small diffusion coefficient causes internal layers to arise
at the positions of the fronts.  The two bound fronts are then divided
into the five regions sketched in Figure~\ref{fig:front}, where $x_L$
and $x_R$ give the positions of the left and right fronts. Within regions
I, III, and V the amplitude of $A$ is constant. Regions II and IV are
regions of rapid change where the dynamics of the fronts are
determined.  The internal layers have a width
$\Delta x={\cal O}(\epsilon)$, implying  $\Delta \tilde{x}=\epsilon^{-2}\Delta x={\cal
O}(\epsilon^{-1})$, which is still large relative to the critical
wavelength of the traveling waves.

We consider the forcing to be small, $\nu = {\cal O}(\epsilon)$, so that the
corresponding amplitude $B$ of the left-traveling wave that is excited by
the forcing is of the
same order in $\epsilon$.   The amplitudes and the parameters  $\mu$ and
$\nu$ are  expanded as \typeout{DON'T FORGET TO INCLUDE THIS
ARRAY}
\bea
A = A_0 + \epsilon A_1 + \cdots, && \hspace{.5in} B = \epsilon B_1 +
\cdots,\label{e:exp1}\\
\mu = \mu_c + \epsilon^2 \mu_2 +\cdots, && \hspace{.5in} \nu = \epsilon \nu_1
+ \cdots.\label{e:exp2}
\eea
where $\mu_c = -3 c^2/16p$ is the value of the control parameter
at which  a single, non-interacting front is stationary. For $\mu
> \mu_c$ the nonlinear convective state invades the basic state.
\typeout{The diffusion scales as $d = \epsilon^2 d_2$ and t} In
the following derivation we go into a reference frame moving with
the group velocity $s$. The positions $x_R$ and $x_L$ of the right
and left fronts evolve then on a slow time scale, $T =
\epsilon^3t$.

Inserting the expansions (\ref{e:exp1}) and (\ref{e:exp2}) into
equations (\ref{e:ampA}) and (\ref{e:ampB}) we obtain at leading
order equations for
$A$ and $B$ in the outer regions I, III, and V,
\bea
0 & = &  \mu_c A_0 +c A_0^3 - pA_0^5,
\label{e:outerA}\\
0 & = &  2 s \partial_x B_1 + \mu_c B_1 +\nu_1 A_0 - g A_0^2 B_1
- u A_0^4 B_1 . \label{e:outerB}
\eea
Solving equation (\ref{e:outerA}) results in $A_0 = 0$ in regions I and V and
$A_0^2 = A_c^2 \equiv 3 c/4p $ in region III. The corresponding solution to
(\ref{e:outerB}) is
\be B_1^j(x) =
    \frac{\nu_1}{\alpha} A_0^j + K^j e^{\frac{\alpha^j}{2s}x},
    \label{e:B1}
\ee
where $j$ corresponds to the regions I, III, and V and $\alpha^j = -\mu_c + g
(A_0^j)^2 + u (A_0^j)^4$.

In the inner regions II and IV, the solutions vary on a fast space scale
$x/\epsilon$ which is captured by introducing the inner coordinates $\eta =
(x-x_L)/\epsilon$ and $\eta = (x_R - x)/\epsilon$, respectively.  The spatial
derivative then transforms as $\partial_x \rightarrow \pm
\partial_\eta/\epsilon$.  The resulting leading-order equations for $A$ and $B$
are
\bea 0 & = &  d_2 \partial_{\eta\eta} A_0 + \mu_c A_0 +c A_0^3 - pA_0^5,
\label{e:innerA0}\\
0 & = &  \pm 2 s \partial_\eta B_1 . \label{e:innerB1}
\eea
From (\ref{e:innerA0}) one obtains the left and right front
solution \be A_0(\eta) = A_c
\sqrt{\frac{1}{2}\left[1+\tanh{\left(\frac{\eta}{\xi}\right)}\right]},
\;\;\;\;\;\; \xi = \frac{4}{c} \sqrt{\frac{p d_2}{3}}, \ee where
$A_c^2$ and $\eta$ are defined above.   Equation (\ref{e:innerB1})
implies that $B_1$ does not depend on the fast space variable
$\eta$, i.e. $B_1 = B_c^{I,IV}$ is constant in regions II and IV.
At ${\cal O}(\epsilon)$, no inhomogeneity arises in the equation
for $A_1$, which is therefore taken to be identically $0$.    But
at ${\cal O}(\epsilon^2)$ the following equation for
$A_2$ is obtained:
\be
{\cal L}A_2 =  \mp\frac{dx_{L,R}}{dT} \partial_\eta A_0
       - \mu_2 A_0 + gB_1^2 A_0 + rA_0^3 B_1^2 - \nu_1 B_1,  \label{e:innerA2}
\ee
where ${\cal L} = d_2 \partial_{\eta\eta} +\mu_c +3cA_0^2 - 5pA_0^4$ is the
linearized operator. It is singular and has the zero-eigenmode
$\partial_\eta A_0$, which leads  to a solvability condition  for equation
(\ref{e:innerA2}). The result of projecting equation (\ref{e:innerA2}) onto the
zero-eigenmode determines the velocity  of the fronts in regions II and IV,

\be
\pm\frac{d x_{L,R}}{dT} =  \left[-2 \mu_c + (2 g + A_c^2 r)
B_c^{II,IV^2}- \frac{4 \nu_1 B_c^{II,IV}}{A_c}\right]\xi  \label{e:frontvel}
\ee
with $B_c^{II,IV}$ yet undetermined.

Since $B$ is generated by $A$, we consider the case where $B=0$
ahead of the pulse  (i.e. for a right-traveling pulse region V where
$A\equiv 0$).  This implies that $K^V = 0$ from   equation
(\ref{e:B1}).  Now matching the inner and outer solutions for $B$
at the left and right positions ($x_L$  and $x_R$) of the fronts,
one obtains the constant values $B_c^{II} =  (\nu_1 A_c/\alpha)
\left[1-\exp{(-\alpha L /2s)}\right]$ and $B_c^{IV} = 0$. The
constants
$K^I$ and $K^{III}$ are nonzero and we see from equation
(\ref{e:B1}) that $B$ grows spatially to the left in region III
and approaches the limiting value $\nu_1 A_c/\alpha$. Then, in
region I where the pulse has passed and $A$ again equals $0$, $B$
decays exponentially to zero.  The individual front velocities are
given by substituting the values of $B_c^{II,IV}$ into the
expressions  given in~(\ref{e:frontvel})
\bea
\frac{d x_{L}}{dT} & = & \left[-2 (\mu - \mu_c)
             + \frac{\nu^2(2 g + A_c^2r)A_c^2}{\alpha^2}
              - \frac{4 \nu_1^2}{\alpha}\right]\xi , \label{e:Lvel} \\
\frac{d x_{R}}{dT} & = & +2 (\mu - \mu_c)\xi.
             \label{e:Rvel}
\eea
Combining these results yields the following equations describing
the evolution of the pulse length $L = x_R - x_L$ and the velocity
of the pulse relative to a frame moving with the group velocity
$s$ in terms of a ``center of mass" coordinate $M = (x_R+x_L)/2$,
\bea
\frac{dL}{dT} & = & k_1(\mu-\mu_c) + k_2 \nu^2 (1-e^{-\frac{\alpha}{2s}L})
               -k_3 \nu^2(1-e^{-\frac{\alpha}{2s}L})^2, \label{e:Ldot} \\
\frac{dM}{dT} & = & \frac{1}{2}\left [-k_2 \nu^2 (1-e^{-\frac{\alpha}{2s}L})
               +k_3 \nu^2(1-e^{-\frac{\alpha}{2s}L})^2\right ],
           \label{e:Mdot}
\eea
where $k_1 = 4 \xi$, $k_2 = 4\xi/\alpha,$ and  $k_3 =
[(2g+A_c^2r)A_c^2 \xi]/\alpha^2$.    While both $k_1 \mbox{ and }
k_2$ are always positive, $k_3$ may be either positive or
negative.  As will be seen in the next section, the sign of $k_3$
determines whether stable pulse solutions may exist.  The
interaction length is given by $2s/\alpha$.

\section{Discussion of Front Equations} \label{sec:discussion}

\begin{figure}[ht]
\centerline{\epsfxsize=2.4in{\epsfbox{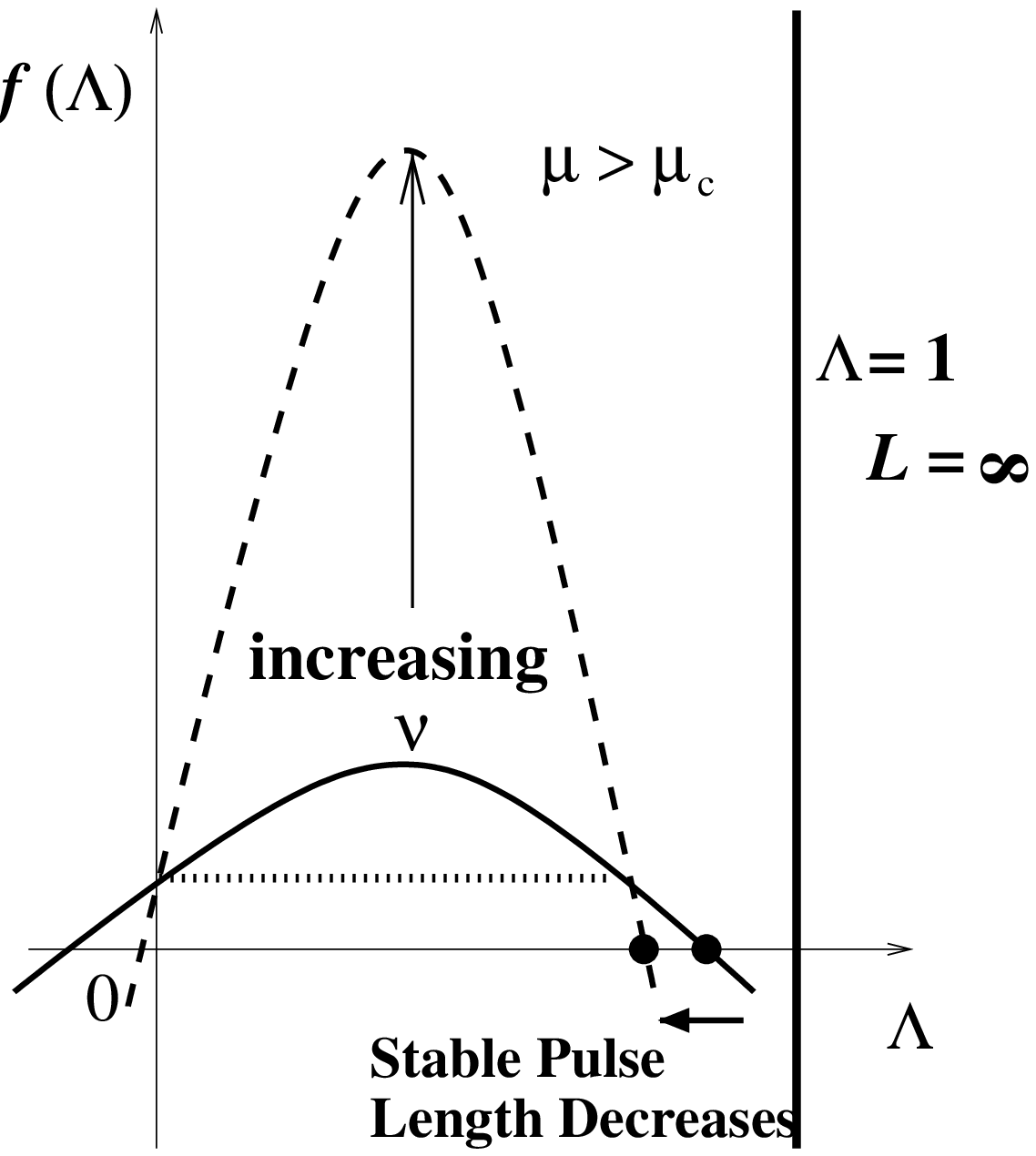}}\hspace{.6in}
            \epsfxsize=2.4in{\epsfbox{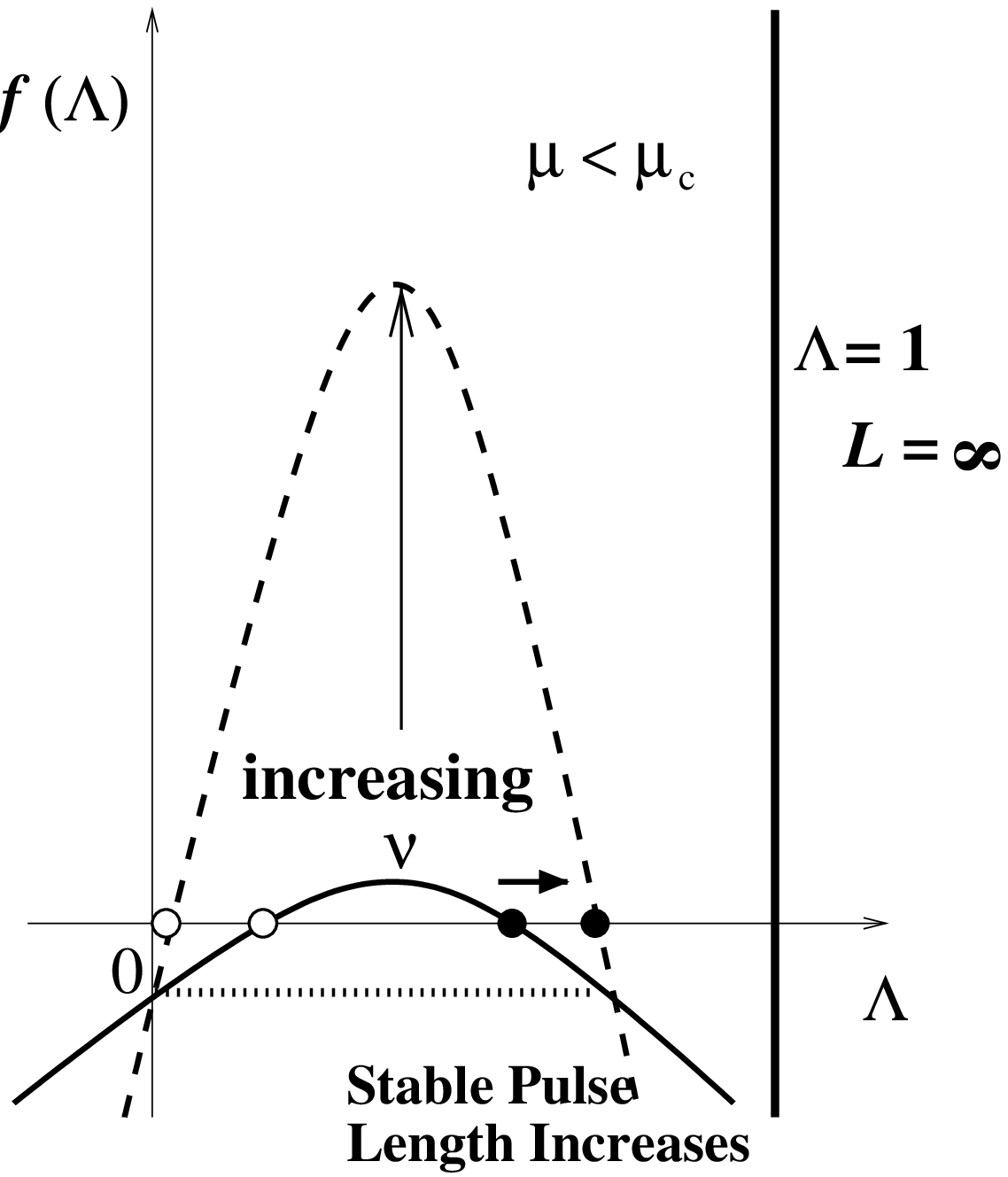}}
}
\caption{Sketch of $f(\Lambda) \equiv k_1(\mu-\mu_c) + k_2\nu^2\Lambda
-k_3\nu^2\Lambda^2$ for $k_3>0$.   Stable (unstable) fixed points
are indicated by solid (open) circles.  Varying forcing changes the steepness
of the parabola and the height of the maximum located at $\Lambda = k_2/2k_3$.
Increasing forcing $\nu$ decreases the stable pulse length for (a) $\mu >
\mu_c$  and increases the stable pulse length for (b) $\mu < \mu_c$.
       }
\label{fig:dtLambda}
\end{figure}

The possible pulse solutions of~(\ref{e:Ldot}) and (\ref{e:Mdot})
are more easily discussed in terms of the quantity $\Lambda \equiv
\left[1-\exp{(-\alpha L /2s)}\right]$ which monotonically
increases from zero to one as $L$ increases from zero to infinity.
Equation~(\ref{e:Ldot}) is then given by
\be
\frac{1}{1-\Lambda}\cdot \frac{d\Lambda}{dT} = f(\Lambda) \equiv
k_1(\mu-\mu_c) + k_2\nu^2\Lambda -k_3\nu^2\Lambda^2.
\label{e:flambda}
\ee
Figure~\ref{fig:dtLambda}  depicts $f(\Lambda)$ for the two cases
$\mu>\mu_c$ and $\mu<\mu_c$.  For $k_3>0$ the parabola opens
downward and the maximum occurs at $\Lambda = k_2/2k_3$,
independent of forcing strength $\nu$. Fixed points $\Lambda_0$ are
indicated where $f(\Lambda) = 0$:
\be
   \Lambda_0 = (1-e^{-\frac{\alpha}{2s}L_0})
   \equiv \frac{k_2 \pm \sqrt{k_2^2 +\frac{4(\mu-\mu_c)k_1 k_3}{\nu^2}}}{2k_3}
\label{e:Lam0}
\ee

Pulses with $\Lambda =
\Lambda_0$ are linearly stable if $f'(\Lambda_0) < 0$. If $k_3 < 0$, the
parabola is opening upward with the minimum at $\Lambda = k_2/2k_3
< 0$ so that $f'(\Lambda) >0$ for all values of $\Lambda
> 0$.  Hence for a stable pulse to exist it is necessary that $k_3>0$ and in this
case corresponds to the upper branch of solutions~(\ref{e:Lam0}).
More precisely, stable pulses exist as long as $k_3>k_2/2$ with their length
diverging for $k_3 \rightarrow k_2/2$. Unless stated otherwise the discussion of the
front interaction will focus on the regime $0<k_2/2k_3<1$ where a
stable branch exists.

From the expression~(\ref{e:Lam0}) and the parabola in
Figure~\ref{fig:dtLambda}, one can easily see the effect of
forcing on the stable pulse length.
As the forcing $\nu$ increases, $f(\Lambda = 0) =
f(\Lambda = k_2/k_3)
= k_1(\mu-\mu_c)$  remain fixed while the parabola steepens (see
Figure~\ref{fig:dtLambda}).  Thus for $\mu
> \mu_c$ the stable pulse length is larger than $L_c$
 corresponding to $\Lambda_c \equiv k_2/k_3$, but approaches
$L_c$ as forcing increases
(Figure~\ref{fig:dtLambda}a). Conversely, for $\mu < \mu_c$ the
pulse length increases to $L_c$ with increasing $\nu$
(Figure~\ref{fig:dtLambda}b).  As forcing $\nu$ decreases for
$\mu >\mu_c$, the pulse length increases and eventually diverges to infinity.
In contrast, for $\mu<\mu_c$  decreasing forcing results in the
pulse length decreasing until the pulse is destroyed in a
saddle-node bifurcation.

\begin{figure}[ht]
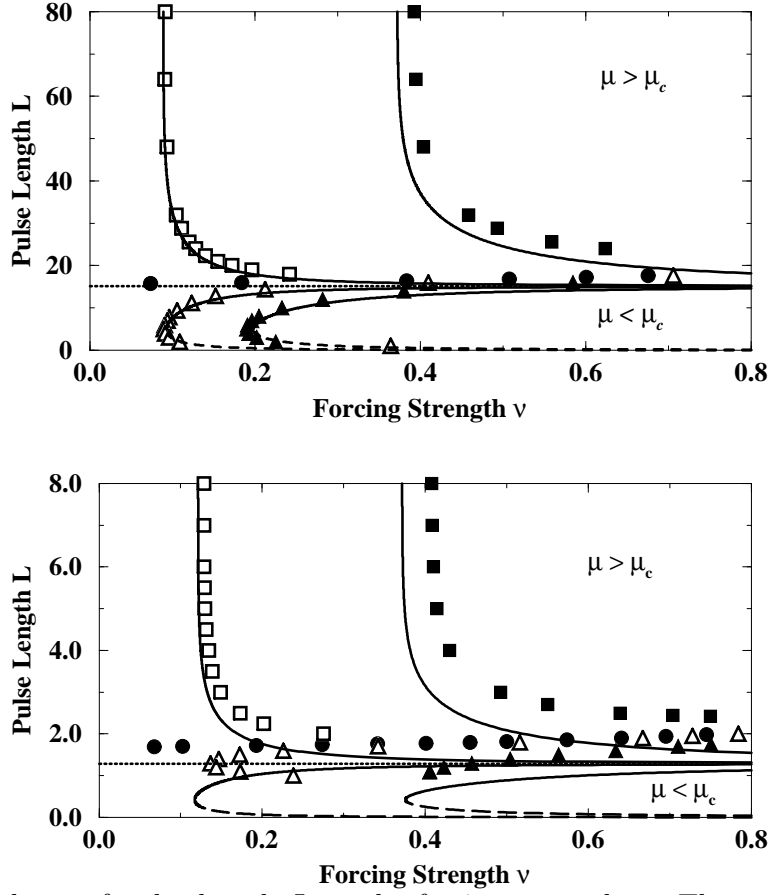

\centerline{\epsfxsize=4.in{\epsfbox{figures/Lvsbs20.bps}}}
\vspace{.2in}
\centerline{\epsfxsize=4.in{\epsfbox{figures/Lvsb.bps}}}
\caption[Dependence of pulse length $L$ on the forcing strength
$\nu$]{Dependence of pulse length $L$ on the forcing strength
$\nu$.  The symbols indicate the numerical results and the curves
are  obtained from equation (\protect{\ref{e:Lam0}}).  The dotted
line corresponds to $\mu = \mu_c$.  The curves above (below) this
line correspond to values of $\mu > \mu_c$ ($\mu<\mu_c$). The
dashed curves indicate  unstable branches. The group velocity $s
= 20 \mbox{ and } 1.7$ in the top and bottom figures, respectively.
$\mu = -1.238$ (solid squares), $-1.2475$ (open squares),
$-1.248075=\mu_c$ (circles), $-1.2485$ (open triangles), and
$-1.250$ (solid triangles). All other parameters are as in
Figure~\protect{\ref{fig:ABsoln}}.
    }
\label{fig:Lvsb}
\end{figure}

Figure~\ref{fig:Lvsb} shows both the analytical
results for the steady state solutions of equation~(\ref{e:Ldot})
and numerical results obtained by integrating the full amplitude
equations~(\ref{e:ampA}) and (\ref{e:ampB}).  A linearized
Crank-Nicholson scheme was used to solve the coupled equations and
the pulse lengths were measured at half-amplitude. The pulse
length is plotted as a function of forcing strength for several
values of the control parameter
$\mu$ confirming the expected behavior. The top figure is for
group velocity $s = 20.0$ and the bottom figure for $s = 1.7$.
Here we see that although the analytical result no longer agrees
quantitatively for smaller
$s$, it still describes the qualitative behavior of the pulse
length as the control parameters are varied.  This dependence on
the group velocity will be discussed at the end of this section.

A numerical control technique has been employed to obtain the
unstable pulse solutions indicated by the long dashed curves in
Figure~\ref{fig:Lvsb}. Since the analysis leads to a single
ordinary differential equation (\ref{e:Ldot}) describing the
evolution of a pulse, the dynamics of the pulse length are
essentially restricted to a one-dimensional manifold. The control
technique is therefore relatively straight-forward.  Pulse lengths
are measured after evolving equations~(\ref{e:ampA}) and
(\ref{e:ampB}) over a short time interval and compared with the
measurements taken at the previous time to obtain the direction
and rate of growth.  This information is then used along with the
desired pulse length to adjust the control parameter accordingly.
This process is repeated until a steady-state pulse with the
specified length is obtained. In Figure~\ref{fig:Lvsb}, the
parameter adjusted to control the length is the forcing strength
$\nu$ while all other parameters remain fixed.  The same technique
is also used for varying the control parameter $\mu$ (cf.
Figures~\ref{fig:hole}b, \ref{fig:holec2.58}, and
\ref{fig:nummoddisp}).

The different regimes can be understood by looking at the individual
interaction terms in equation (\ref{e:Ldot}) and their origins from
equation (\ref{e:ampA}).  The first term in (\ref{e:Ldot}) is a measure of
how far the control parameter $\mu$ is from the critical value $\mu_c$
where, in the absence of forcing, an isolated front is stationary.  It
provides a ``pressure" which is directed outward for $\mu>\mu_c$
and which has to be balanced by the interaction terms.  The second
and third terms describe the interaction between fronts due to forcing.
The second term arises from the linear coupling between $A$ and
$B$ introduced by the forcing in equation (\ref{e:ampA}) through
which $B$ excites $A$. It therefore enhances the invasion of the
nonlinear state into the linear state, which corresponds to a repulsive
interaction between the leading and the trailing front.  The third term
stems from the nonlinear coupling which, when $k_3>0$, suppresses
$A$ and therefore weakens the invasion, implying attraction between
the fronts.

\begin{figure}[ht]
\centerline{\epsfxsize=3.5in{\epsfbox{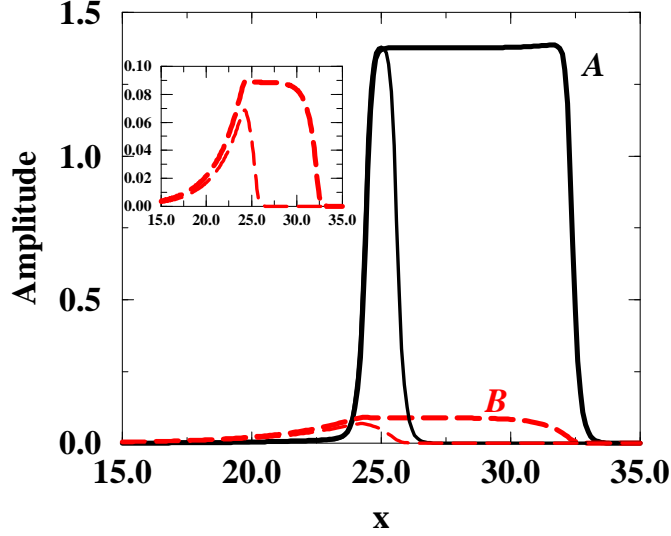}}
}
\caption{Numerically obtained long and short stable pulse
solutions, with
$\mu> \mu_c$ and $\mu<\mu_c$, respectively.  The inset zooms in on
the amplitude $B$ only.  The parameters are $\nu = 0.2534, s=1.7,
d =0.05, c = 2.58, p = 1.0, g = 1.4, r = 4.0$, and $\mu = -1.244$
($\mu = -1.250$) for the long (short) pulse. For these parameters
$\mu_c = -1.248075$.
       }
\label{fig:ABsoln}
\end{figure}

\begin{figure}[ht]
\centerline{\epsfxsize=4.0in{\epsfbox{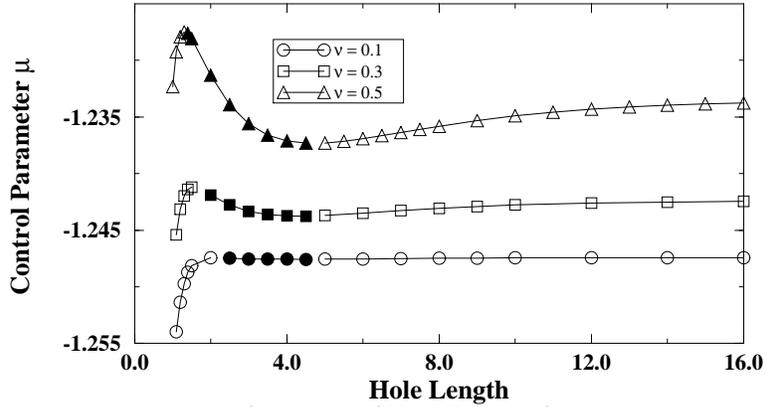}}}
\caption[Control parameter $\mu$ as a function of hole
length]{Control parameter $\mu$ as a function of hole length for
various values of $\nu$.  For $c = 2.58$, $k_3$ is in the regime
where stable holes exist only for $\mu>\mu_c$.  Solid (open)
symbols indicate stable (unstable) solutions.  All other
parameters are as in Figure~{\protect{\ref{fig:ABsoln}}} .
    }
\label{fig:holec2.58}
\end{figure}

The effect of forcing on a pulse depends on the distance between
the fronts. Since the pulse is traveling to the right, the
amplitude $B$ is growing spatially to the left. For short pulses,
therefore, $B$ remains small and the linear coupling term
dominates the interaction implying a repulsive interaction.  Thus, for
$\mu<\mu_c$ the inward ``pressure" can be balanced by a repulsive interaction
if the pulses are sufficiently short.  With increasing pulse lengths, $B$
reaches larger values at the trailing front and the nonlinear
coupling term gains importance. Thus, in contrast to
many other localization mechanisms, the forcing can induce an
{\it attractive} interaction that grows with distance. It is able to balance the outward
``pressure" for $\mu>\mu_c$.  In this regime the pulses become
shorter with increased forcing.  Figure~\ref{fig:ABsoln} shows two
stable pulse solutions obtained by numerically integrating
equations (\ref{e:ampA}) and (\ref{e:ampB}).  The control
parameter $\mu = -1.240$ for the longer pulse and $\mu=-1.250$ for
the  shorter pulse.  All other parameters are the same for both
pulses. Note that the forcing $\nu$ is the same for both
pulses, yet the amplitude
$B$ has not saturated for the shorter pulse.

\begin{figure}[ht]
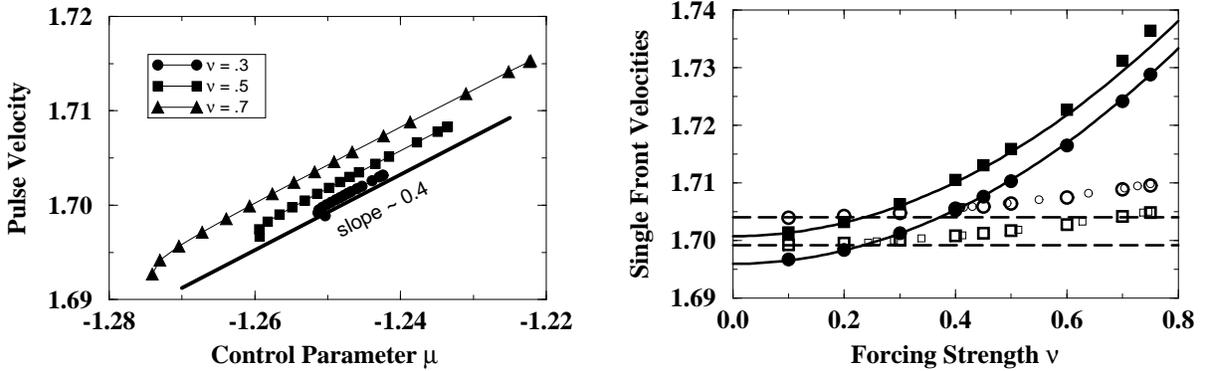

\centerline{\epsfxsize=3.0in{\epsfbox{figures/vvsa.bps}}\hspace{.2in}
        \epsfxsize=3.0in{\epsfbox{figures/vLRpulsevsb.bps}}
        }
\caption{(a) Pulse velocity as a function of $\mu$ for forcing $\nu = 0.3,
0.5$, and $0.7$.  The line shown is obtained from the analytical result
$v_{pulse} = k_1(\mu-\mu_c)/2 +s$.  (b) Velocity dependence on forcing $\nu$.
The solid and open symbols indicate the numerically obtained
velocities for the trailing and leading fronts, respectively. The
curves give the analytical results.  Squares are for $\mu =
-1.250$ and circles for $\mu = -1.238$.  The small symbols show
the corresponding numerically obtained pulse velocities.  All
other parameters are as in Figure~{\protect{\ref{fig:ABsoln}}}.
    }
\label{fig:vel}
\end{figure}

\begin{figure}[ht]
\centerline{\epsfxsize=3.0in{\epsfbox{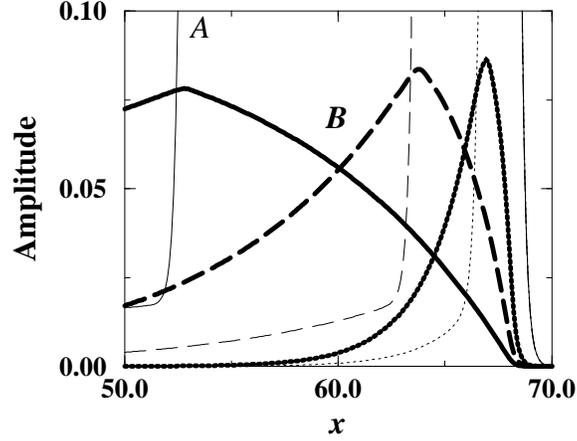}}
}
\caption{Amplitude $B$ of pulse solutions when $s=1.25, 5.0, 20.0$
indicated by the thick dotted, dashed, and solid lines,
respectively.  The amplitudes $A$ are indicated by thin lines and
the leading front in $A$ is indistinguishable between the
different velocities. $\mu = -1.2485, \nu = 0.29$ and all other
parameters as in Figure~{\protect{\ref{fig:ABsoln}}}.
    }
\label{fig:solnvarys}
\end{figure}

\begin{figure}[ht]
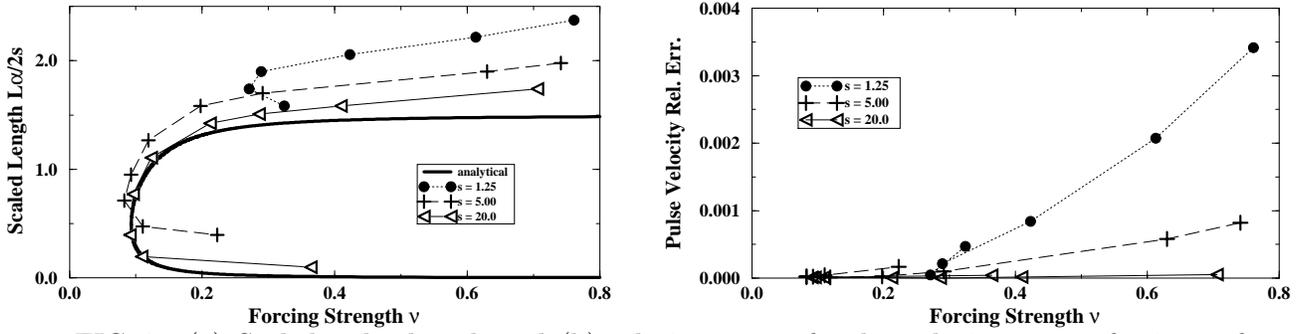

\centerline{\epsfxsize=3.2in{\epsfbox{figures/Lalpvsba12485.bps}}
\hspace{.2in}\epsfxsize=3.3in{\epsfbox{figures/velabsrelerra12485.bps}}
}
\caption{(a) Scaled pulse length  and (b) relative error of pulse
velocity versus forcing $\nu$ for $s=1.25, 5.0, 20.0$. $\mu =
-1.2485$ and all other parameters as in
Figure~{\protect{\ref{fig:ABsoln}}}.
        }
\label{fig:Lvvarys}
\end{figure}

From equations (\ref{e:Ldot}) and (\ref{e:Mdot}) we see that when
$dL/dT\equiv 0$ for a steady pulse $L_0$, then $dM/dT =
k_1(\mu-\mu_c)/2$.  Thus the pulse velocity in the moving frame is
given by the invasion speed and depends on $\mu$.  By contrast, in
dispersively stable pulses the velocity depends on the control
parameter only through the nonlinear gradient terms
\cite{MaNe90,HaPo91}.  Strikingly, in the asymptotic calculation
leading to (\ref{e:Mdot}) the pulse velocity does not depend on
the forcing $\nu$.  Figure~\ref{fig:vel}a shows that the linear
dependence on $\mu$ is in good agreement with this analytical
result, but that the velocity does also depend on forcing $\nu$.
We suggest that the discrepancy may be explained by the effect of
the excited counter-propagating wave on the leading front at
$x_R$.  Since in the weakly nonlinear regime diffusion is much
weaker than advection (for group velocities of ${\cal O}(1)$),
$B_c = 0$ in region
IV, which implies that forcing has no affect on the velocity of
the leading front (equation \ref{e:frontvel}).  However, non-zero
$B$ within this inner region will slightly alter the velocity of
the leading front and consequently that of the pulse as a function
of forcing. This will also impact the length of the pulse. For the
parameters used in Figure~\ref{fig:ABsoln}, note that $B$ varies
slowly in region II whereas it varies almost as fast as $A$ in
region IV suggesting that the missing contribution is more
prevalent at the leading front than at the trailing one.
Figure~\ref{fig:vel}b shows single front velocities for both the
leading and trailing front.  The curves are the analytical results
obtained from (\ref{e:frontvel}) for single fronts ($L \rightarrow
\infty$) and the symbols indicate numerical results.  As expected,
the velocity of the trailing front is well described by our
analysis, but the velocity of the leading front does depend
slightly on the forcing.  The leading-front velocity then
dictates the velocity of the pulse. For larger group velocity $s$,
the interaction length
$2s/\alpha$ increases and the separation of the fast and slow spatial scales at
the positions of the fronts becomes more distinct, as shown in
Figure~\ref{fig:solnvarys}.  Thus we expect and the numerical
results confirm (Figure~\ref{fig:Lvvarys}) that the agreement with
analytical results (\ref{e:Ldot}) and (\ref{e:Mdot}) for pulse
length and velocity improves with increasing group velocity $s$.
For small forcing $\nu$ the relative error of the velocity is on
the order of the numerical accuracy.  We note that the qualitative
behavior of the pulse length is consistent even for smaller group
velocity.

\section{Holes and Multiple Pulses}
\label{sec:holes}

The analysis from section \ref{sec:derivation} can be applied to other
front configurations such as hole states and multiple pulses.   A hole
consists of a localized region in which the amplitude of the pattern is
very small and  which is surrounded by the nonzero traveling-wave
amplitude. In contrast to the hole-type solutions found in the
Ginzburg-Landau equation \cite{LeFa97,He98}, which are
qualitatively different objects than pulses, the holes to be discussed
here  are similar to pulses in that they are also compound objects
made of two fronts connecting the basic state with the  nonlinear
state. As with the pulses, the central equation for their description is
an evolution equation for the length  $L_h$ of the hole. It is given by
\be
\frac{dL_h}{dT} = -k_1(\mu-\mu_c) - k_2 \nu^2 (1+e^{-\frac{|\mu_c|}{2s}L_h})
               +k_3 \nu^2(1+e^{-\frac{|\mu_c|}{s}L_h}), \label{e:Lhdot}
\ee
where $k_1, k_2$, and $k_3$ are defined as before and a stable hole may be
possible when $k_3 >0$.  Note that the interaction length for holes is now
given by $2s/|\mu_c|$ which is longer than the interaction length  $2s/\alpha$
for pulses.  Solving $dL_h/dT = 0$ again yields two possible solutions
$L_h^\pm$, with the longer hole unstable and the shorter hole stable.
In contrast with the pulses, stable holes thus exist only over a
finite range of shorter lengths.  The RHS is quadratic in
$\Lambda_h = \exp(-|\mu_c| L_h/2s)$, describing a parabola opening
upward.  Although varying forcing now results not only in
stretching the parabola, but also in shifting it vertically, the
variations in hole length again depends on the sign of the first
term in (\ref{e:Lhdot}).  When
$\mu<\mu_c$, increasing forcing causes stable holes to get shorter, while
for $\mu>\mu_c$ the stable hole length will increase.  Although the tendency of
a hole to grow or shrink with increased forcing depends only on the sign of
$\mu-\mu_c$, the limiting behavior is different depending on the relative sizes
of $k_2$ and $k_3$.   If $k_3<k_2$ then holes exist only for $\mu<\mu_c$ and as
$\nu$ increases, the hole length goes to zero.  If
$k_3>k_2/2(\sqrt{2}-1)\approx 1.2k_2$, holes exist only for $\mu>\mu_c$ and as
$\nu$ increases the stable hole length grows and eventually disappears in a
saddle-node bifurcation with the unstable hole.  Finally, for the intermediate
values of $k_3$, holes exist for values of $\mu$ both above and below $\mu_c$.
In this case, as $\nu\rightarrow \infty$, the hole length approaches the
limiting length determined when $\mu=\mu_c$.

\begin{figure}[ht]
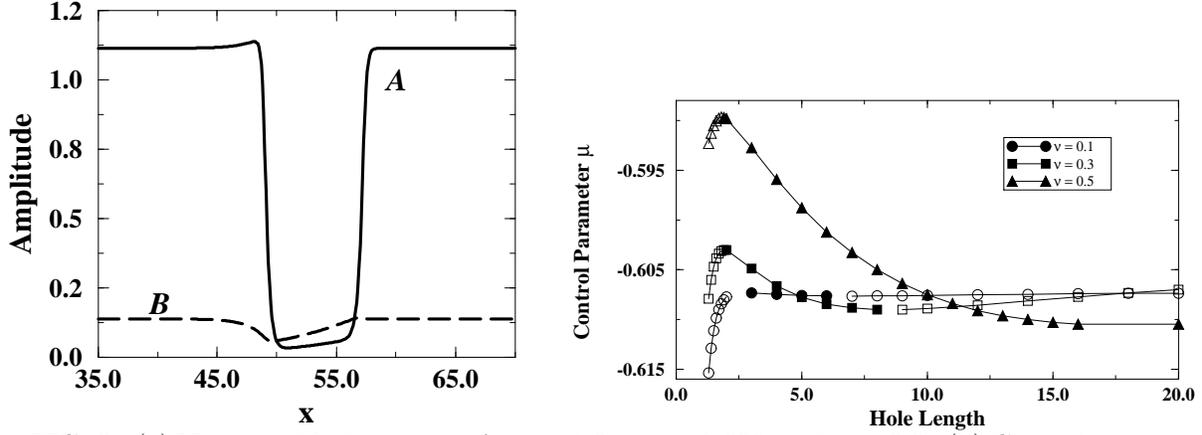

\centerline{\epsfxsize=2.7in{\epsfbox{figures/holeb.bps}}\hspace{.2in}
            \epsfxsize=3.3in{\epsfbox{figures/avsL.bps}}
}
\caption{(a) Numerical hole solution for $c = 1.8, \mu = -0.609$,
and $\nu = 0.3$.  (b) Control parameter $\mu$ as a function of
hole length for various values of $\nu$.  For $c = 1.8$, $k_3$ is
in the regime where holes exist for both $\mu>\mu_c$ and
$\mu<\mu_c$.  Solid (open) symbols refer to stable (unstable)
solutions.  All other parameters are as in
Figure~{\protect{\ref{fig:ABsoln}}} .
    }
\label{fig:hole}
\end{figure}

Figure~\ref{fig:hole}a,b shows a stable numerical hole solution and
the control parameter $\mu$ as a function of the hole length for
different values of the forcing.  Stable and unstable solutions
are indicated by solid and open symbols, respectively.  Again, the
unstable holes are obtained by means of a numerical control
technique.  First we note that according to (\ref{e:Lhdot}), the
minimum of these curves should all be at the same value of $L_h$,
but our numerical results show the minimum shifted to the right as
$\nu$ increases.  The analysis requires that $A$ vanish in the
hole region, but the presence of $B$ in this region actually
generates small nonzero $A$, which in turn generates $B$. Hence,
the actual value of $B$ at the trailing front is greater than
predicted.  This suppresses $A$ there and the trailing front slows
down, leading to longer pulses and a shift of the minimum of the
curves to the right as forcing $\nu$ increases.  If the parameters
are such that the control parameter $\mu$ can be taken smaller,
the basic state is more strongly damped.  Hence, within the hole
region $A$ is smaller and the shift of the curves is less
pronounced as seen in Figure~\ref{fig:holec2.58}.  Here $c=2.58$
so that $\mu_c \sim -1.248$ as compared to $c=1.8$ in
Figure~\ref{fig:hole} where $\mu_c \sim -0.608$.

\begin{figure}[ht]
\centerline{\epsfxsize=3.5in{\epsfbox{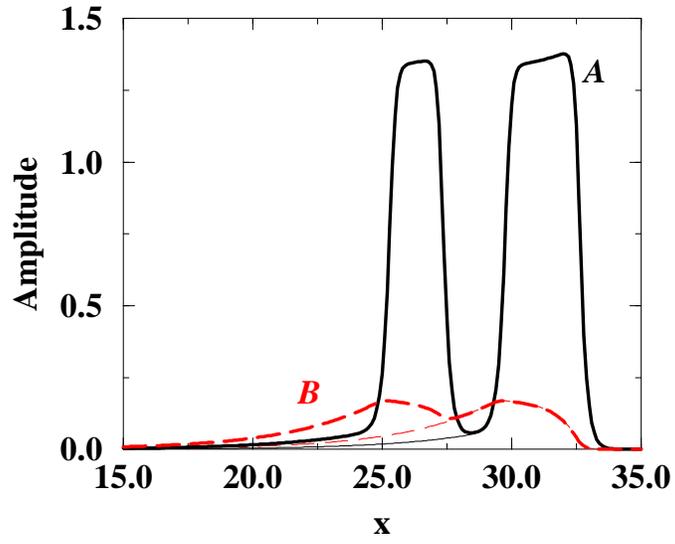}}
}
\caption{Two-pulse solution (thick lines) superimposed over a
single-pulse solution (thin lines) for $\mu=-1.238, \nu = 0.5$,
and all other parameters as in
Figure~{\protect{\ref{fig:ABsoln}}}.
    }
\label{fig:twopulse}
\end{figure}

Arrays of multiple fronts can be combined to form multiple
pulses.  For a two-pulse configuration, there are four inner front
regions which match to five outer regions resulting in evolution
equations for the distance $L_2$ between the pulses as well as the widths
$L_1$ and $L_3$ of the leading and the trailing  pulse, respectively.
This is to be contrasted with the description of multi-pulse solutions in
the strongly dispersive case (without forcing), where the pulse widths can be
adiabiatically eliminated in favor of the
the distance and the phase difference between the two individual
pulses \cite{AkAn97}. The three evolution equations for $L_{1,2,3}$ are given by
\begin{eqnarray}
\frac{dL_1}{dT} & = &  k_1(\mu-\mu_c) + k_2\nu^2\Lambda_1
                            -k_3\nu^2\Lambda_1^2, \label{e:dtL1}\\
\frac{dL_2}{dT} & = & -k_1(\mu-\mu_c)
                            - k_2 \nu^2 \Lambda_1(1+\Lambda_2)
               +k_3 \nu^2 \Lambda_1^2(1+\Lambda_2^2), \label{e:dtL2}\\
\frac{dL_3}{dT} & = &  k_1(\mu-\mu_c) + k_2\nu^2 \left\{\Lambda_1\Lambda_2
                   + \left[1+\Lambda_3(\Lambda_1\Lambda_2 - 1) \right] \right\}
           \nonumber \\
                   && \;\;\;\;\;\;\;\;-k_3 \nu^2 \left\{\Lambda_1^2\Lambda_2^2
                   + \left[1+\Lambda_3(\Lambda_1\Lambda_2 - 1) \right]^2 \right\},
           \label{e:dtL3}
\end{eqnarray}
where the $\Lambda_i$ are defined in terms of $L_i$ as follows:
$\Lambda_1 = 1-\exp(-\alpha L_1/2s), \Lambda_2 = \exp(\mu_c
L_2/2s)$, and $\Lambda_3 = \exp(-\alpha L_3/2s)$.  Since equation
(\ref{e:dtL1}) for $L_1$ is the same as equation (\ref{e:Ldot}) which
describes a single pulse, the leading pulse length is unaffected by
the trailing pulse. But the trailing pulse length depends on both the
length of the leading pulse and the distance between the pulses and
is typically shorter than the leading pulse. Solving equations
(\ref{e:dtL1}) and (\ref{e:dtL3}) for the fixed points $\Lambda_{10},
\Lambda_{20}, \Lambda_{30}$, it is found that $\Lambda_{10}\cdot
\Lambda_{20} = k_2/k_3$ and $\Lambda_{30} =
k_3\cdot(1-\Lambda_{10}/(k_3-k_2)$.  Since $L_1$ increases with
increasing $\Lambda_1$, and $L_2$ and $L_3$ increase with
decreasing $\Lambda_2$ and $\Lambda_3$, this suggests that all
three lengths will  either increase or decrease as the control
parameters $\mu$ and $\nu$ are varied. The eigenvalues $\sigma_i$
obtained from linearizing  equations (\ref{e:dtL1}) and (\ref{e:dtL3})
indicate that $\sigma_1 = \sigma_3$ and a two-pulse solution may be
stable  when $k_3$ is positive.  A stable two-pulse solution is
therefore expected to exist and be stable whenever a single pulse
exists.  However, since the trailing pulse is narrower than the leading
pulse, it may become too short and collapse - as it merges in a
saddle-node bifurcation with the shorter unstable pulse - for
parameter values where the single pulse is still stable. Numerically,
we observe that as the control parameter $\mu$ is decreased all
three lengths get shorter and the trailing pulse eventually collapses to
zero. Figure~\ref{fig:twopulse} shows a stable two-pulse solution
superimposed over a single pulse solution confirming that the leading
pulse length is unaffected by the trailing pulse.

\section{Dispersion Effects}\label{sec:disp}

Waves generally have both linear and nonlinear dispersion.  Hence,
the coefficients and the amplitudes in equations (\ref{e:ampA})
and (\ref{e:ampB}) are in general complex.  In
\cite{MaNe90,HaPo91}, it is shown that for weak dispersion the
interaction between fronts leads to the following type of
evolution equation for the length of a pulse:

\be \frac{dL}{dt} = k_1(\mu-\hat{\mu}_c) - k_4 e^{-L/\xi} + \frac{k_5}{L}.
\label{e:dispLdot} \ee
The coefficients $k_1$ and $k_4$ are positive and
contain only the real part of the original  coefficients. The
first term is similar to equation (\ref{e:Ldot}) where
$\hat{\mu}_c$ is the value of the control parameter at which a
single, isolated front is stationary.  Due to the dispersive terms,
$\hat{\mu_c}$ includes a correction compared to $\mu_c$.
 The second term arises as a
result of the overlap of the fronts in the convective amplitude.
The term involving $k_5$ contains the imaginary parts of the
coefficients, so that this term represents the interaction due to
dispersion.  When $k_5 > 0$, dispersion provides a repulsive
interaction and can lead to the existence of stable pulses.  When
$\mu<\hat{\mu}_c$ two pulse solutions exist with the longer one
being stable.

We now consider the combined effects of forcing and
dispersion. Of particular interest is the question
whether periodic forcing can stabilize or
destabilize pulses obtained in the purely dispersive regime.  The
task of deriving front equations including both dispersion and
forcing seems formidable.  It is suggested and validated by the
numerical simulations below that the relevant aspects of both
features may be modeled by simply adding the two contributions. This
leads to an equation of the following form for pulses of length
$L$:
\be
  \frac{dL}{dT} = k_1(\mu-\hat{\mu}_c) + k_2 \nu^2 (1-e^{-\frac{\alpha}{2s}L})
               -k_3 \nu^2(1-e^{-\frac{\alpha}{2s}L})^2
           - k_4 e^{-L/\xi} + \frac{k_5}{L}. \label{e:mdLdot}
\ee

\begin{figure}[ht]
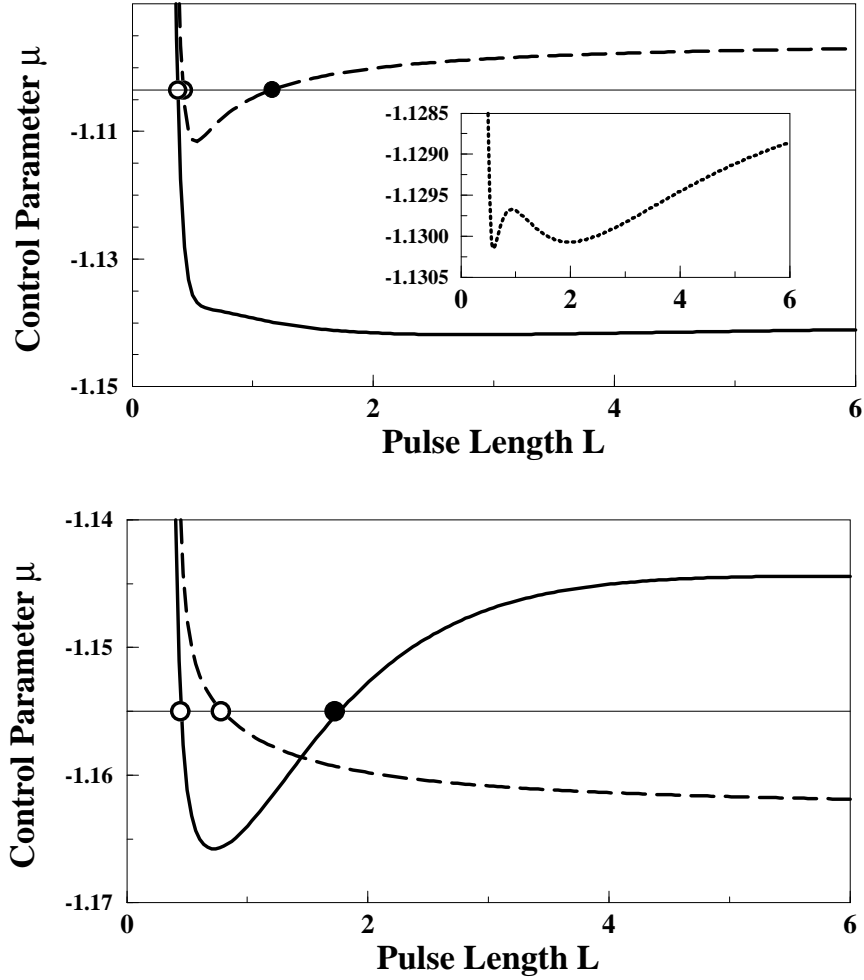

\centerline{\epsfxsize=4.5in{\epsfbox{figures/modkilldisp.bps}}}
\vspace{.2in}
\centerline{\epsfxsize=4.5in{\epsfbox{figures/modstabdisp.bps}}
}
\caption[Pulse lengths for dispersion and added forcing]{Control
parameter $\mu$ vs. steady-state pulse lengths of equation
(\protect\ref{e:mdLdot}) for dispersion only, $\nu = 0$ (dashed
curve), and added forcing, $\nu = 0.5$ (solid curve).
The solid circles indicate stable solutions (positive slope)
and the open circles unstable ones (negative slope).
(a) For $k_5>0$ and $k_2/2k_3>1$ forcing can eventually destroy a stable
pulse, $d = 0.01 +0.0i, r = 0.0$.  The inset shows the four
branches possible for an intermediate value of forcing $\nu =
0.425$. (b)  For $k_5<0$ and
$k_2/2k_3<1$ increased forcing can lead to stable pulse solutions, $d =
0.01-0.01i, r = 4.0$. All other parameters are the same: $s=1.7, c
= 2.45 + 0.2i, p = 1.0, g = 1.4$.
        }
\label{fig:dispmod}
\end{figure}

If both dispersion and forcing independently result in a stable pulse,
their combined contribution  merely enhances the stability of the
pulse.  The case of competing contributions is more interesting.
Figure~\ref{fig:dispmod} is a plot of the control parameter $\mu$ as a
function of steady-state pulse lengths $L$ from
equation~(\ref{e:mdLdot}). The dashed curve is for dispersion in the
absence of forcing ($\nu=0$), whereas the solid and dotted curves
indicate the addition of forcing. For a given value of the control
parameter indicated by the thin horizontal line stable equilibrium
solutions (positive slope) are given by solid circles while open circles indicate unstable
solutions (negative slope). Plotted are both cases, a) when weak
dispersion leads to a stable pulse, $k_5>0$, and b) when it does
not, $k_5<0$. If forcing does not stabilize a pulse ($k_2/2k_3>1$),
then increasing forcing pushes the stable branch of solutions
downward. There is a competition between the two interactions and
four solution branches may exist (Figure~\ref{fig:dispmod}a inset).
Eventually, for sufficiently large forcing $\nu$ the stable pulse
disappears at infinity, leaving only the unstable pulse solution.
Similarly, when dispersion does not stabilize the pulse, increasing
forcing with $k_2/2k_3<1$ can eventually lead to a stable pulse
branch (Figure~\ref{fig:dispmod}b).

For large $L$, equation~(\ref{e:mdLdot}) is dominated by the dispersive
interaction, which decays only like $1/L$,
suggesting that the original, dispersive behavior is recovered for
large lengths $L$.  If $k_2/2k_3 > 1$ the competition can lead to two
additional pulses, one unstable and one stable, for intermediate values of
forcing $\nu$ (cf. inset of Fig.~\ref{fig:dispmod}a). For larger forcing, only
one unstable pulse and one long, stable pulse remain.  Note that for fixed
$\mu$ and increasing $\nu$ the stable pulse will still disappear at infinity
since the branch is being pushed downward.  If the forcing generates a stable
pulse, the $1/L$-behavior of the dispersive interaction suggests the existence
of a long, third, unstable pulse (Fig.~\ref{fig:dispmod}b).  Note that
according to the analysis of the dispersive interaction in \cite{MaNe90}, which
was confirmed numerically in \cite{RiRa95}, the power-law behavior continues
only up to a maximal length $L_{max}$, beyond which the dispersive interaction
decays rapidly.  The maximal length $L_{max}$ decreases with increasing
dispersion.  If forcing dominates the interaction for lengths up to $L_{max}$,
then the long-length dispersively dominated behavior is no longer expected for
either case depicted in Figure~\ref{fig:dispmod}. But it should be present for
weaker forcing or shorter interaction length.

\begin{figure}[ht]
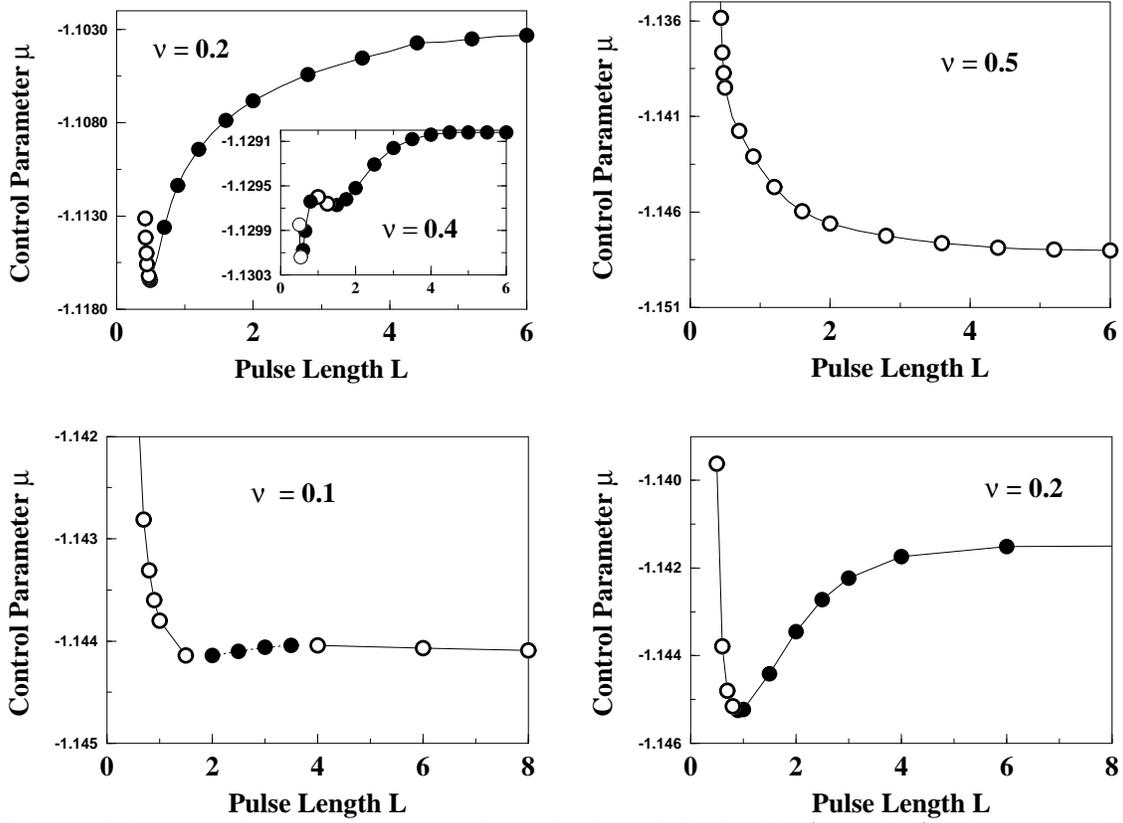

\centerline{
  \epsfxsize=2.8in{\epsfbox{figures/modkillb2_4c245.bps}}\hspace{.2in}
  \epsfxsize=2.8in{\epsfbox{figures/modkillb5c245.bps}}
}\vspace{.2in}
\centerline{
  \epsfxsize=2.8in{\epsfbox{figures/modstabb1c245.bps}}\hspace{.2in}
  \epsfxsize=2.8in{\epsfbox{figures/modstabb2c245.bps}}
}
\caption[Pulse lengths when forcing and dispersive contributions
compete]{The control parameter $\mu$ vs. the pulse length L.  Stable
(unstable) solutions indicated by solid (open) symbols.  Top row
(a)-(b) Forcing destroys a dispersively stable pulse ($d = 0.01 +
0.0i$ and $r = 0.0$). Forcing strength $\nu = 0.2$ and $\nu = 0.4$
(inset) in (a), and $\nu = 0.5$ in (b). Bottom row (c)-(d) Forcing
creates a stable branch ($d = 0.01 - 0.005 i$ and $r = 4.0$). Forcing
strength $\nu = 0.1$ in (c) $\nu = 0.2$ in (d).   All other parameters
are the same: $s=1.7, c = 2.45 + 0.2i, p = 1.0, g = 1.4$.         }
\label{fig:nummoddisp}
\end{figure}

Figures~\ref{fig:nummoddisp} show numerical results that confirm
the expected pulse behavior based on equation~(\ref{e:mdLdot}).
Plotted are the numerically obtained pulse lengths as a function
of the control parameter $\mu$. Stable pulse solutions are
indicated by solid symbols and unstable solutions by open
symbols. Figures~\ref{fig:nummoddisp}a,b show that increased
forcing may destroy a dispersively stable pulse. For forcing
strength
$\nu=0.2$, the forcing is not sufficient to eliminate the stable
pulse and both solution branches are still present.  The inset
shows an intermediate value of $\nu=0.4$ where four solution
branches exist and Figure~\ref{fig:nummoddisp}b shows the pulse
lengths for stronger forcing $\nu=0.5$, where only the unstable
branch remains. Figures~\ref{fig:nummoddisp}c,d depict the
creation of a stable pulse branch by increasing forcing strength.
When $\nu=0.1$ forcing leads to a stable branch, but for longer
lengths the dispersion dominates and a third, unstable pulse
exists. However, for $\nu = 0.2$ the forcing succeeds in
dominating the interaction beyond $L_{max}$ so that the branch
remains stable.

\section{Conclusions}
\label{sec:conclusion}

In this paper we have investigated the effect of external resonant
forcing on the interaction of fronts connecting the stable basic
with a stable nonlinear traveling-wave state that arises
 in a subcritical bifurcation.
Localized structures were described analytically as bound pairs of
fronts.  The temporal forcing excites the oppositely traveling
wave and provides an additional mode that is sufficient to
localize structures. Since the forcing constitutes an externally controlled
parameter, pulses of tunable length can be obtained via this new
localization mechanism. Forcing stabilizes pulses through either a
repulsive or an {\em attractive} interaction between the fronts
depending on the pulse length.  This is in contrast to other
localization mechanisms in which stable pulses arise only for a
repulsive interaction \cite{MaNe90,HaPo91,HeRi95}.  Multiple
pulses with different, but fixed, lengths and holes are also
obtained. In addition, the combined effect of temporal forcing and
dispersion has been investigated.  With the inclusion of weak
dispersion, the interaction between fronts can be described qualitatively
 by a single equation combining the two interaction
terms~(\ref{e:mdLdot}).  It was found that in the dispersive
regime forcing can lead to the creation of new pulses or the
destruction of pulses depending on system parameters.  The
competition between the two interactions determines the number and
stability of pulses observed.

In the regimes investigated here no complex dynamics of the individual pulses
have been found. It is known, however, that both dispersively stabilized pulses
as well as pulses stabilized by an advected mode can undergo transitions to
chaotic dynamics (e.g. \cite{DeBr94,HeRi97,SoAk00}).

Localized traveling waves in the absence of resonant forcing have
been observed experimentally in particular in binary-mixture convection and
electroconvection in nematic liquid crystals. Depending on
parameters, the localization of pulses in binary-mixture
convection is understood to be due to dispersion and to the
coupling to a slowly decaying, advected concentration mode. Since
the advection depends on the direction of propagation of the wave,
an interesting question is how the counterpropagating wave excited
by a resonant forcing will affect the pulses that are stabilized
by the concentration mode and how the two localization mechanisms
interact. The  origin of the localization  of the worms in
electroconvection is still being investigated. A Ginzburg-Landau
model that includes a coupling to a weakly damped mode similar to
that in the case of binary-mixture convection has been proposed
\cite{RiGr98}.  By probing the response of the worms to temporal
forcing,  insight may be gained into the relevance of an advected
mode in the worms.

We would like to acknowledge discussions with J. Vi\~{n}als. This work has been
supported by the Engineering Research Program of the Office of Basic
Engineering Science at the Department of Energy under grant DE-FG02-92ER14303
and the National Science Foundation under grant DMS-9804673.

\newpage

\newpage

\bibliography{journal}

\end{document}